\documentclass[aps,prx,onecolumn,superscriptaddress]{revtex4-2}
\usepackage[utf8]{inputenc}
\usepackage{amsmath}
\usepackage{hyperref}
\usepackage{mathdots}
\usepackage{dsfont}
\usepackage{amsfonts}
\usepackage{amssymb}
\usepackage{subcaption,graphicx}
\usepackage{filecontents}
\usepackage{tikz}
\usetikzlibrary{quantikz}
\usepackage{multirow}
\usepackage{xcolor}
\usepackage{bbold}

\begin{document}

\title{Quantum computing for simulation of fluid dynamics}

\author{Claudio Sanavio}
\email{claudio.sanavio@iit.it}
\affiliation{Fondazione Itituto Italiano di Tecnologia, Italy}

\author{Sauro Succi}
\affiliation{Fondazione Itituto Italiano di Tecnologia, Italy}
\affiliation{Mechanical engineering Department, University College London, UK}

\begin{abstract}

We present a pedagogical introduction to a series of quantum
computing algorithms for the simulation of classical fluids,
with special emphasis on the Carleman-Lattice Boltzmann method.
\end{abstract}

\maketitle

\section{Introduction}

Quantum computing has been vigorously pursued in the recent years, mostly in view
of the spectacular speedup it may provide as compared to the performance of 
electronic computers, for a number of scientific applications \cite{QC}.  
Quantum computing can be traced back to Richard Feynman's 
epoch-making paper, in which he observed that physics "ain't classical", hence 
it ought to be simulated on quantum computers \cite{FEY}.
Following Feynman's observations (with due credit to previous investigators
such as T. Toffoli and E. Fredkin), early theoretical work on quantum computing 
was performed in the 1980s, e.g. Deutsch's work on the link between quantum 
theory, universal quantum computers and the Church–Turing principle\cite{Deutsch1985}. 
Then, with the publication of Shor's algorithm for integer factoring and Grover's 
search algorithm in the middle of the 1990s, the research area gathered significant 
momentum in terms of theoretical work and quantum computing hardware as well.
The research area of quantum computing has continued to grow ever since  
\cite{Preskill2018,IBM,Bravyi2022}. 

\begin{figure}
\centering
\includegraphics[scale=0.3]{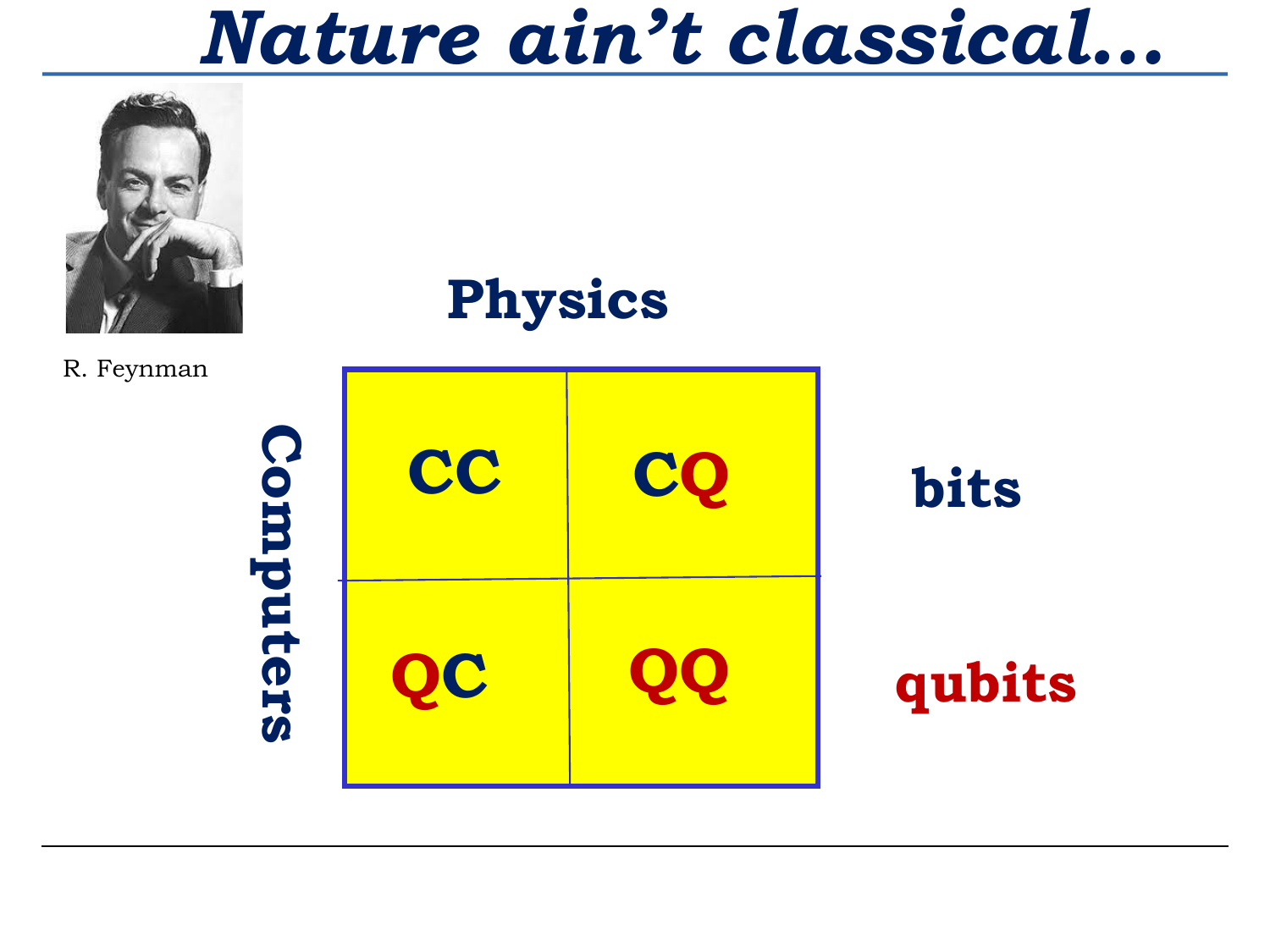}
\caption{The four quadrants in the Physics-Computing plane. 
The CC and CQ quadrants are the mainstay of current scientific computing.
QQ is the quadrant invoked by Feynman, and QC is the quadrant relevant
to this paper.
}
\label{fig_CCQQ}
\end{figure}

In terms of applications, the simulation of quantum 
many-body systems has received special attention, due to its 
scientific and industrial relevance, as well as due to its close 
link with quantum hardware, meaning by this the possibility of
mapping quantum hamiltonian directly into native quantum gates. 
In this paper, we shall focus on a much less beaten track, namely
the use of quantum computers for the simulation of classical fluids
\footnote{Despite their early appearance, we shall not discuss the so-called type-II 
quantum computers \cite{YEPEZ1}, since they do not appear
to comply with universal quantum computing}.
To this purpose, it is convenient to refer to the physics-computing 
plane defined by the following four-quadrants:
%\begin{itemize}
CC: {\it Classical computing for Classical physics};
CQ: {\it Classical computing for Quantum physics};
QC: {\it Quantum computing for Classical physics};
QQ: {\it Quantum computing for Quantum physics}.
%\end{itemize}
%
as shown in Figure \ref{fig_CCQQ} (taken from \cite{QC_EPL}).

% ---------------------
% ---------------------
Feynman's observation pertains to the CQ sector shown in 
Figure \ref{fig_CCQQ}, where one is often confronted with exponential 
complexity barriers associated with the phase-space of quantum
many-body problems \cite{PVC,GG}. 
The basic idea is that such exponential barriers can be handled by
the corresponding exponential capacity of qubit representations 
offered by the QQ quadrant.
In this Chapter, we shall focus on the opposite off-diagonal QC quadrant, where
the power of quantum computing might be brought to the fruition 
of hard computational problems in classical physics.
In particular, we shall focus on computational fluid dynamics, which
presents a ceaseless quest for better and more efficient algorithms and
computers, mostly on account of the problem of fluid turbulence \cite{PRAMA,QC_EPL}.
The physics of fluids presents two general features which are not shared by
quantum physics: {\it nonlinearity} and {\it dissipation}.
Both set a major challenge to quantum computing, which draws much of its 
power from the superposition principle and on the unitary and hamiltonian
structure of quantum mechanics. 
In the following we shall present a few potential strategies to deal with both
issues above.

\section{Challenges facing quantum computational fluid dynamics (QCFD)}
As mentioned in the introduction, realising the potential of quantum computing means 
leveraging distinctive features of quantum mechanics that, by definition, are not available 
on classical computers.  However, it also follows that it is precisely these specific features that 
expose major challenges in realising simulations with a quantum advantage.

The main quantum mechanical concepts spawning the potential benefit of quantum 
algorithms are {\it quantum superposition} and {\it quantum parallelism}. 
The quantum state in an $Q$-qubit coherent register can be described by the 
Schr\"odinger wave function, defined by $2^Q$ complex amplitudes for $2^Q$ states in superposition. 
The square of each of these amplitudes defines the probability of finding the system 
in the corresponding state after {\it quantum measurement}.
By encoding classical data in terms of these amplitudes an exponential saving in 
storage can be achieved when the number of qubits is compared to required number of classical bits. 

Let us illustrate the idea for the specific case of turbulent flow simulation. 
Turbulence features a $Re^3$ complexity, where the Reynolds number $Re$ represents 
the relative strength of convection (nonlinearity) versus dissipation. 
Most real life problems feature Reynolds numbers in the many-millions; for instance 
an airplane features $Re \sim 10^8$, implying $O(10^{24})$ floating-point operations per
simulation. This is basically the best one can afford on a nearly-ideal Exascale classical
computer. The simulation of regional atmospheric circulation flows takes us at least another
two decades above in the Reynolds number, hence totally out of reach for any
foreseeable classical computer \cite{WEATHER,PALMER}.
On the other hand, the minimum number of qubits $Q$ required 
to represent $Re^3$ complexity can be roughly estimated as
$2^Q = Re^3$, namely:
\begin{equation}
Q =3 Log_2 Re \sim 10 log_{10} Re
\end{equation} 
This simple estimate shows that $80$ qubits match the requirement of full-scale
airplane design, while $O(100)$ qubits would enable regional atmospheric 
models \cite{WEATHER} or nuclear fusion applications \cite{QC_Fusion}.
However, several key challenges stand in the way of this task.
%
%\begin{itemize}

{\it First, Quantum measurement:}  Extracting classical information 
collapses the quantum wave vector into a single eigenstate (a house of cards). 
Hence, to get classical values for each of the amplitudes, multiple 
realisations of the quantum state vector are needed with an associated set of measurements.

{\it Second, Conditional operations:} In the quantum circuit model, the 'classical' information 
is not available to the quantum gate operations performed in the circuit. 
Specifically, gate operations can be conditional on the state of one or 
more control qubits, while specifying gate operations conditional on 
one or more of the complex amplitudes defining the wave function is impossible. 
Therefore, when classical data are encoded in terms of amplitudes, a 
rotation angle in a quantum gate operation, this information is required at the 
time the circuit is compiled.  This angle cannot be changed at run-time as a function 
of 'classical' data encoded in quantum amplitudes.

{\it Third, Time marching:} In an algorithm involving multiple iterations or time-steps, the overhead 
associated with quantum measurements used to extract classical data and re-initialization 
of the quantum state for a next iteration, scale quadratically with the grid size
\cite{STAN} and consequently they severely tax the quantum CFD efficiency.
It should be observed that the well-known HHL\cite{HHL} 
algorithm for linear system solution, assumes that the input and output data are
encoded in terms of quantum amplitudes without including
the cost of setting up the quantum state and extracting the classical solution.

{\it Fourth, Circuit depth:}
With the exception of quantum measurement operations, quantum mechanical operators 
are unitary, linear and reversible, hence they can be directly implemented 
on a series of unitary quantum gates (quantum circuit model).
Usually, in order to quantify the computational cost, producers and specialists 
take track of the number of two-qubit gates employed in the circuit. 
In fact, two-qubit gates, such as the CNOT gate, come in the hardware with an error that is 
several orders of magnitude (depending on the platform) larger than single-qubit gates errors. 
We know from the DiVincenzo theorem~\cite{BARENCO} that the number of two-qubits gates needed 
to implement a generic $Q$ qubits unitary operation scales as $4^Q$, thus voiding the exponential 
advantage of the embedding process. 
In order to sidestep this issue, the unitary operation that we want to perform in our circuit 
has to be \textit{combinatorially block diagonal}~\cite{BERRY}. 
Hence, this provides a stringent constraint on the possible algorithms that can
actually be implemented to simulate classical fluids.

If we further assume that the velocity field is represented in terms 
of {\it amplitude encoding}\cite{Brassard2002}, i.e. the components of velocity vector 
at all grid points are represented in terms of the amplitudes defining the 
wavefunction, then the {\it no-cloning theorem} prevents the use of 
(temporary) copies of any of these amplitudes. 
So, evaluation of $u^2$ or $u \frac{\partial u}{\partial x}$ cannot be performed 
by storing a temporary copy $temp=u$, to perform the computation 
of the value of $u^2$ as $u \times temp$. 
Also, for data encoded in terms of the complex amplitudes of the Schr\"odinger wavefunction, there 
is a need for this state vector to have a unit norm, since these amplitudes represent 
probabilities of states. 
This means that for an operator attempting to compute the squares of these 
complex amplitudes, the resulting state vector loses unitarity. 
So, even without the no-cloning theorem complicating such a step, this points to a 
further problem with computing non-linear terms. 
This loss of unit norm of the quantum state vector represents an example of 
a {\it non-unitary operation}, typically associated with a corresponding loss of information.
A similar argument runs for orthogonality of the eigenstates, since a nonlinear
propagator rotates the state by an angle which depends on the state itself.

{\it Fifth, Dissipation:} Dissipation breaks hermicity of the quantum propagator.
This can be recovered by augmenting the system with its hermitian 
conjugate, so that the doubled system is hermitian by construction.
One can dispense with doubling degrees of freedom by 
representing the non-unitary update as a weighted sum of unitaries.
This introduces a failure probability which needs to be handled
probabilistically, i.e. by executing the update conditional on the
successful measurement of an ancilla qubit.  

To summarize, dealing with non-linear and dissipative terms raises 
major extra challenges for Quantum Computational Fluid Dynamics (QCFD)
besides the ones commonly encountered for the simulation of quantum systems.

\section{Hybrid quantum/classical approaches}

The challenges sketched above have resulted 
in the fact that most of the existing work in QCFD is based on 
a {\it hybrid quantum/classical approach}, with the quantum processor 
performing computations for which efficient quantum algorithms exist, while 
the output is then passed on to classical hardware to perform further 
computational tasks not (yet) amenable to quantum algorithms. 

\begin{figure}
\centering
\includegraphics[scale=0.3]{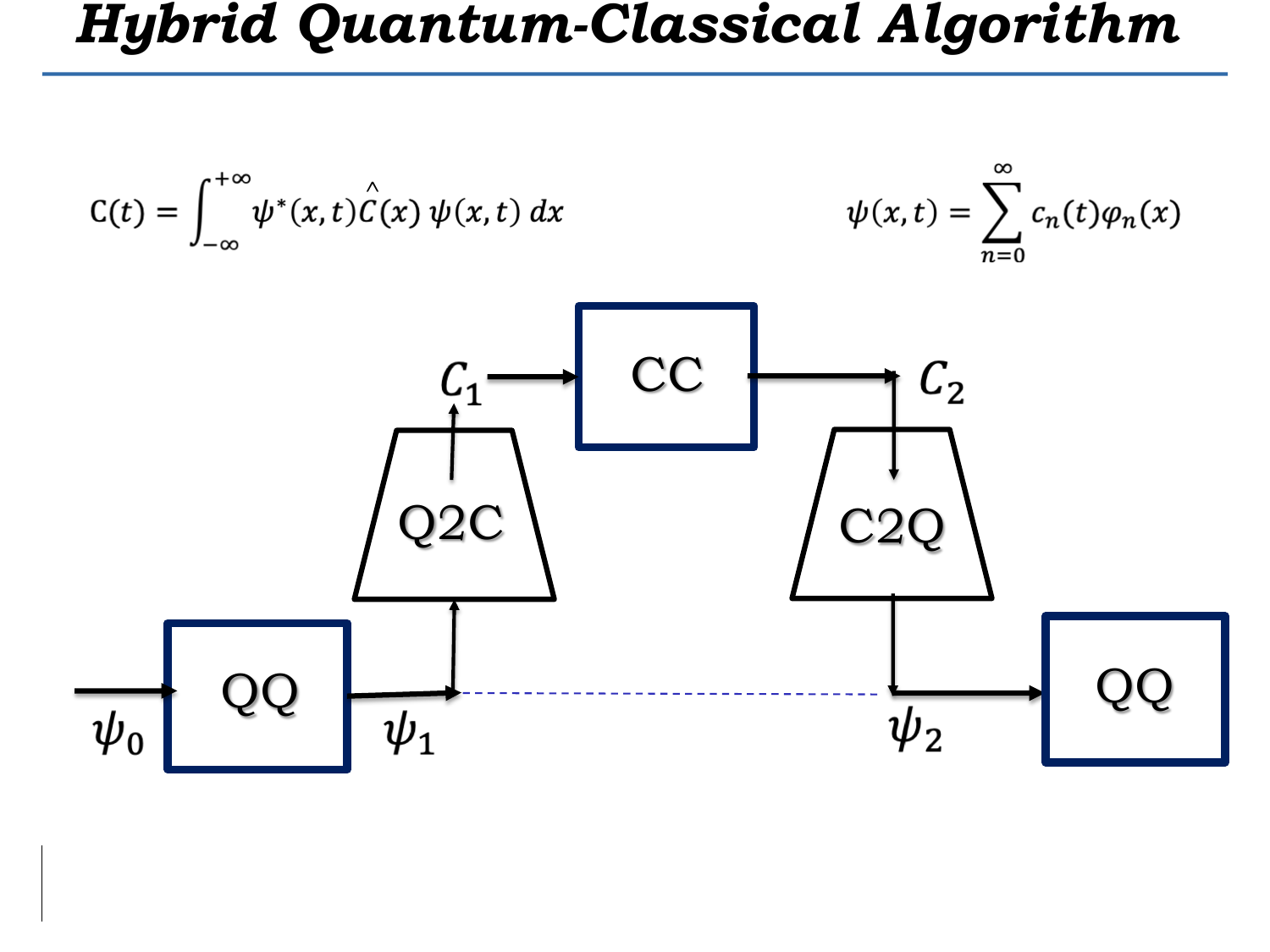}
\caption{Sketch of a hybrid quantum-classical algorithm. 
The illustration shows the steps involved in a single time step or single 
iteration, including preparation of the subsequent step/iteration.}
\label{fig_HybridQC}
\end{figure}

Figure \ref{fig_HybridQC} illustrates the idea.
The quantum state $\psi_0$ is advanced to 
the next quantum state $\psi_1$ via a QQ algorithm. 
The quantum state $\psi_1$ is then measured to generate 
classical observables $C_1$ which are advanced to $C_2$ by a CC algorithm.
The classical observables $C_2$ are then used to reconstruct the quantum 
state $\psi_2$, ready for the next QQ step. 
The Q2C conversion shown in Figure \ref{fig_HybridQC} involves quantum measurements and 
requires repeated measurements over a statistical sample of quantum states, since
none of them can be reused. The C2Q reconstruction requires the preparation
of all the quantum eigenstates, hence a full reset of the quantum circuits from scratch. 
Both operations impose a substantial computational burden
on the hybrid algorithm, typically scaling exponentially with the number of qubits. 

Examples of previous works using the {\it hybrid quantum/classical approach} include 
the works of Steijl \cite{RENE2019}, Gaitan \cite{GAITAN} and Budinski \cite{Budinski2022}. 
The algorithm presented by Gaitan uses Kacewiz's quantum amplitude estimation ODE 
algorithm \cite{KACEW} as applied to the set of nonlinear ODE's resulting from standard 
discretization of the Navier-Stokes equations. 
As shown, for certain 'non-smooth' problems (illustrated using the quasi-1D flow in 
converging-diverging duct with normal shock wave), the complexity analysis shows potential 
for exponential speed-up, so that the challenges associated with hybrid quantum/classical computing 
can potentially be overcome. 
For the linear advection-diffusion equation, Budinski\cite{Budinski2021} presented a quantum 
algorithm based on the Lattice Boltzmann method\cite{OUP1,LB}. 
The algorithm can perform multiple successive time steps with no need 
for quantum measurements and re-initialization of qubit register between the time 
steps if suitable re-scaling of solution vector is applied to deal with non-unitarity. 
In the quantum circuit model, the fact that velocity field is unchanged between successive 
time steps implies that the operator implementing $u \frac{\partial u }{\partial x}$ 
can be re-used in multiple time steps. 
Budinski\cite{Budinski2022} then extended the work to Navier-Stokes equations in 
streamfunction-vorticity formulation. Then, velocity-field updates during each time step means 
that quantum-circuit implementation of convection terms cannot be re-used during multiple 
time steps and that the classical value of $u$ (as well as other flow field data) is needed 
to define quantum circuit implementation for the next time step. 
This shows that it is the non-linearity that forces the use of a 
hybrid quantum/classical approach, similar to previous work where 
Navier-Stokes equations were discretized, e.g. the hybrid quantum/classical 
algorithms for fluid simulations based on quantum-Poisson solvers\cite{RENE2019}.

In summary, key challenges for the {\it hybrid quantum/classical approach} are:
i) Cost and complexity of (repeated) measurements;
ii) Statistical noise due to sampling of the quantum solution;
iii) Cost and complexity of (repeated) re-initialization.

The development of more efficient (re-)initialization techniques appears to be a key
factor for the future success of quantum computing for fluids 

% ---------------------
% ---------------------

\section{Quantum fluid dynamics strategies}

For the sake of concreteness, let us refer to the
Navier-Stokes equations for time-dependent incompressible flows:

\label{sect_nonlinear}
\begin{eqnarray}
&&\frac{\partial {\bf u}}{\partial t}
+ {\bf u}\cdot\frac{\partial {\bf u}}{\partial {\bf x}}=
-\frac{\partial P}{\partial {\bf x}} + \nu \Delta {\bf u}
\label{eq_NS_mom}\\
&&\nabla \cdot {\bf u} = 0
\label{eq_NS_mass}
\end{eqnarray}
where ${\bf u}$ is the velocity vector, $P$ is the pressure, defined for location 
${\bf x}$ as function of time $t$. The kinematic viscosity is 
defined by $\nu$ and density has been conventionally set to a unit constant value. 
Equation (\ref{eq_NS_mass}) enforces mass conservation, while Equation (\ref{eq_NS_mom}) is 
based on momentum conservation in each coordinate direction.
Equations (\ref{eq_NS_mom}) and (\ref{eq_NS_mass}) highlight that it is the convection term that represents 
the non-linearity, i.e. the second term on the left-hand side of Equation (\ref{eq_NS_mom}). 
Writing the Navier-Stokes equations in non-dimensional 
form, such that ${\bf x}$ and ${\bf u}$ are scaled by reference length $L_{ref}$ and $U_{ref}$, respectively, it 
follows that in Equation (\ref{eq_NS_mom}) the term $\nu$ is replaced by $1/Re$, where Reynolds 
number $Re=U_{ref} L_{ref}/\nu$. 
For Stokes flow, i.e. with Reynolds approaching $0$, nonlinear 
terms are vanishingly small, but still not to be neglected since they 
are responsible for non trivial long-range correlations especially 
important in biological flows \cite{RMP}. 
For high Reynolds number (turbulent) flows, obviously the nonlinear terms play the leading role.

Different strategies have been adopted to simulate fluid dynamics 
on quantum computer, mostly involving hybrid algorithms.
Since they have been reviewed in the recent Perspective \cite{QC_EPL}, here we simply list
them, directing the reader to the original literature:
\subsection{Nonlinear quantum ODE solvers}

This approach consists in treating the discretized Navier-Stokes equations as a set
of nonlinear ODE's and advance them in time by developing bespoken quantum nonlinear
ODE solvers.  This approach has been pioneered by Gaitan \cite{GAITAN} and demonstrated
for the case of a Laval nozzle on grids with $O(30-60)$ grid points over $1400$ time steps.
The critical issue, though, is that the concrete implementation of
the quantum oracle is left to a classical computer.   

\subsection{Nonlinear variational quantum eigenvalue solvers}

it has recently been proposed that variational quantum eigenvalue 
(VQE) solvers, a major tool of the QQ sector, might be extended
to the fluid equations as well \cite{CFDVQE}. 
The basic idea is to use quantum computers for the efficient
generation of variational eigenfunctions by suitable circuit
parametrization. The associated energy functional is then minimized
through a classical procedure.
The appeal of these methods is the possibility to import algorithmic know-how
from quantum mechanical applications, especially quantum chemistry.
To date, however, there is no evidence that the procedure can meet the standards
of accuracy required by computational fluid dynamics.

%In \cite{CFDVQE} the authors propose to use a similar technique for the Navier-Stokes
%equations, i.e construct variational trial functions and minimize the associated
%(dissipative) functional via a classical search algorithm.
%Formally, step 1 consists in expressing the flow field in variational form
%\begin{equation}
%\label{VARU}
%{\bf u}(x,t;\lambda) = \sum_n {\bf u}_n(t) \phi_n(x;\lambda)
%\end{equation}
%where $\phi_n$ is a suitable set of basis functions parametrically 
%dependent on the a set of variational parameters $\lambda$.
%Such variational parameters are then fixed by minimizing the energy 
%dissipation functional $D(\lambda) = \nu \int (\nabla u(x;\lambda))^2 dx$
%where the integral runs over the entire volume occupied by the fluid.
%The appeal of this idea is twofold: first, it may borrow a lot of QQ know-how, second 
%it bypasses the issue of quantum time marching.     
%We are not aware of any practical implementation of the idea, but best guess is that
%they will become available soon.

\subsection{Functional approach} 
It is long known that any nonlinear problem can be mapped into a 
linear one in a space of higher dimensions, a technique
also known as Carleman embedding or Carleman linearization \cite{CARLE}.
The Carleman embedding trades nonlinearity for infinite dimensions and nonlocality. 
An alternative procedure to recover linearity is to take a probabilistic approach and
formulate a functional kinetic (Liouville) equation for the 
the probability distribution function (PDF) of the fluid velocity field.
Formally
\cite{DSFD22}:
\begin{equation}
\partial_t P[{\bf u}] + \frac{\delta}{\delta {\bf u}} 
(f({\bf u}) P[ {\bf u}])=0
\label{LIU}
\end{equation}
where $P=P[{\bf u}]$ is the functional PDF and $\dot {\bf u} = f({\bf u})$ is the 
equation of motion (hence $f({\bf u})$ is a nonlinear and nonlocal  operator in function space). 
This is linear by construction, regardless of the nonlinearity of the dynamics.
Once the fluid equations are discretized on a grid with $G$ lattice nodes, the Liouville 
distribution $P_G$ becomes a $3G$-variate PDF $P_G({\bf u}_1 \dots {\bf u}_G)$
which lives in a $O(3G)$-dimensional space.
For large-scale flow applications, $G$ can reach values up to 
multi-billions, hence fully under the "curse of dimensionality". 

Fortunately, for most practical purposes, low order marginals prove
often capable to retain the essential physical information.
Marginalization is known to introduce coupling terms between 
different marginals, thus raising the need for appropriate closures expressing
the high order marginals as functionals of the low order ones.
This is a classical topic in non-equilibrium statistical physics, which 
may draw significant benefits from modern developments in 
tensor networks theory \cite{TN,TN2}. 

The idea can be taken one level further by noting that
Eq.~\eqref{LIU} is linear, but still not unitary. 
In \cite{GIANNAKIS} the authors presented a method to relate the  classical 
evolution and the unitary evolution of a quantum system, making use of the 
Carleman-Koopman embedding, which transforms the Liouville equation into an
equivalent many-body Schr\"odinger equation.
 
This is formally very appealing, but still subject to the dimensional curse; even
in quantum chemistry nobody knows how to handle molecules with billions atoms
with ab-initio methods.
In addition, \cite{GIANNAKIS} shows that for chaotic systems the spectrum of 
the hamiltonian contains continuous modes that are pretty hard to tame.

Always in the spirit of functional methods, in \cite{Jin2023}, the authors develop 
another form of Koopman-von Neumann approach based on the level-set method as
applied to classical nonlinear  field theories \cite{Jin2023}. 
The formalism is applied to hyperbolic PDE's and Hamilton-Jacobi equations,
but its extension to the Navier-Stokes equations is addressed only 
marginally, making it difficult to draw firm conclusions.

\subsection{Carleman embedding}  
Introducing indices $m$ and $n$ to denote interacting neighbors, and using Einstein summation, 
the fluid equations can be recast in terms of a linear operator 
$\mathcal{L}$ and quadratic operator $\mathcal{Q}$, as follows,
\begin{eqnarray}
\frac{d u_{l}}{dt} = \mathcal{L}_{lm} u_{m} 
+ Re \; \mathcal{Q}_{lmn} u_{m} u_{n}
\label{eq_NS_ODE2}
\end{eqnarray}
The pressure term was removed for simplicity, although this all but a minor item, since
pressure-flow-stress coupling is going to affect the structure of the Carleman matrix. 

At this stage, it should be noted that in the discretization of the velocity derivatives 
at lattice point $l$, the values of the velocity at one or more neighboring lattice points is used. 
This means that when the Carleman linearization is used, it introduces a 
new variable $U_{lm} = u_l u_m$ to formally 
generate a linear system. Marching the system of equations forward in time produces
an ever growing hierarchy in which the Carleman variables at level $k$ involve the Carleman 
variables at the next level $k+1$. 
Note that at each level we are faced with a tensor of rank $k$, 
with $O(G \kappa^{k-1})$ independent 
components, $\kappa$ being the sparsity of the $\mathcal{Q}$ matrix. 

Furthermore the tensors occupy a neighborhood of the original field
whose diameter grows linearly with the Carleman level.
This shows that there is a major prize to pay to uplift the 
nonlinearity of the fluid equations, both in terms of 
increasing dimensionality and loss of locality.
Nevertheless, in \cite{CHILDS} the authors  
present an algorithm based on Carleman linearization along with the 
use of a quantum linear system solution for the solution of the
one-dimensional Burgers equation. 
The presented algorithm shows a polylog scaling with the number 
of grid points, i.e. an exponential improvement over
classical approach. However, for a given time span $T$, the algorithm 
shows a complexity $T^2 Poly(logT)$, i.e. a significantly 
increased time-complexity compared to classical case. 
The authors simulate shock formation in a one-dimensional Burgers flow, with 
$16$ grid points over $4000$ time-steps, with a fourth order Carleman truncation.
They reach Reynolds numbers up to $40$, an order of magnitude
larger than predicted by the theoretical no-go analysis, a welcome discrepancy
which begs for further analysis.

The key question then is: is there a maximum Reynolds number above which
the Carleman procedure fails to converge? Also, when it does converge, 
how many Carleman levels are required to reach the desired accuracy at
a given Reynolds number?

The answer may well depend on the chosen representation of the fluid 
equations and in the following we turn to a promising candidate in this 
respect, namely the Lattice Boltzmann method \cite{OUP1,OUP2}. 

\section{Carleman Lattice Boltzmann}

\begin{figure}
\includegraphics[scale=0.45]{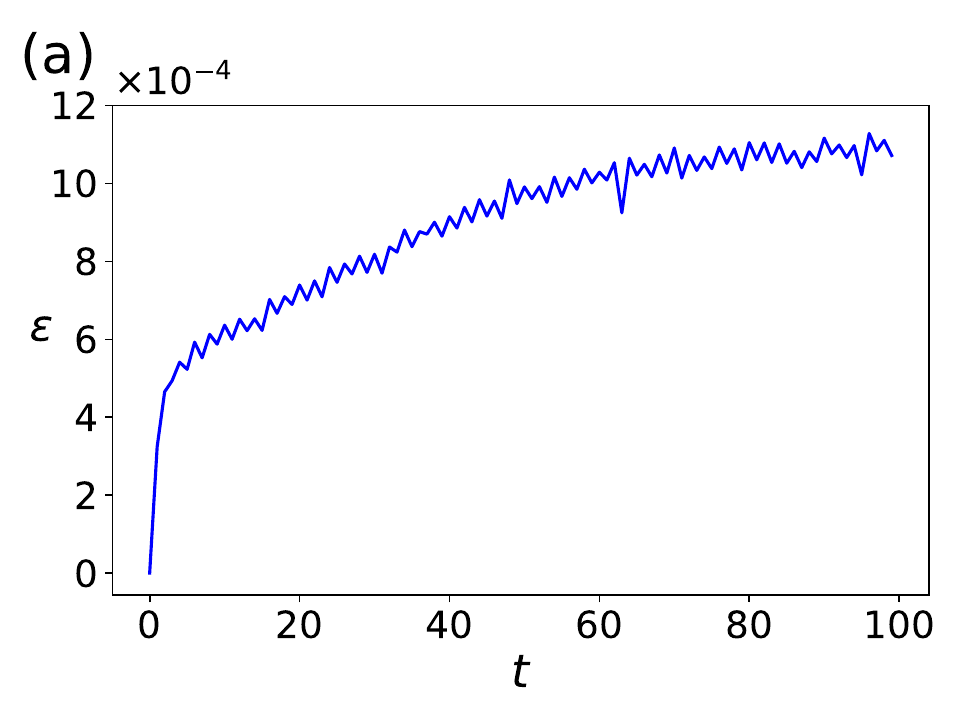}
\includegraphics[scale=0.45]{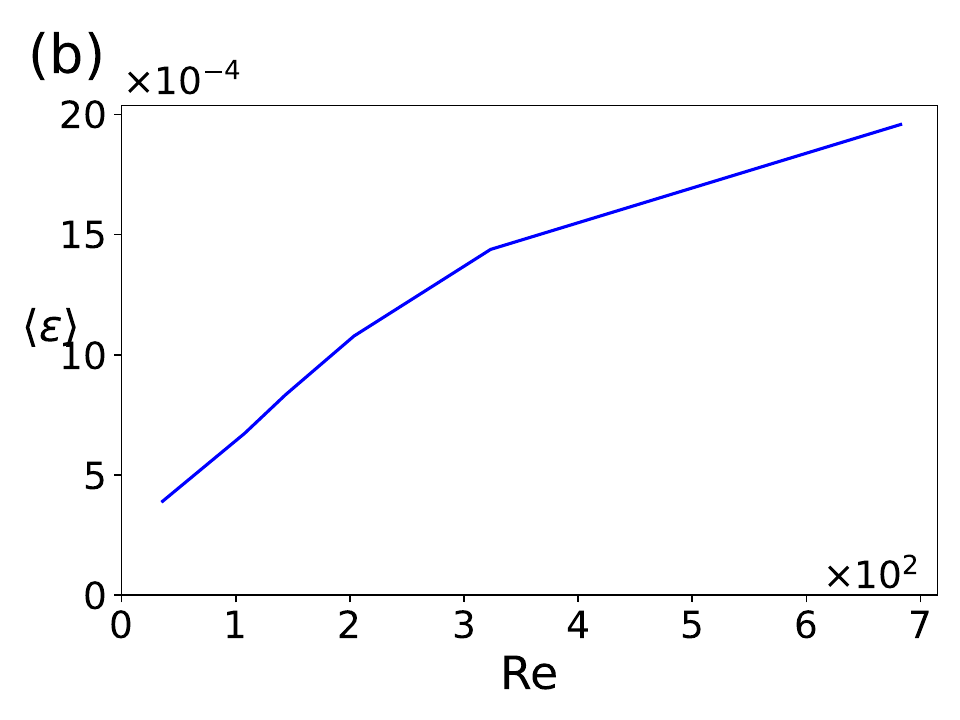}
\caption{(a) The deviation $\epsilon$ in time of the second order Carleman
solution from the original LB solution for the case of a Kolmogorov-like
flow at $Re \sim 10$, on a $32 \times 32$ grid. (b) The mean deviation $\langle\epsilon\rangle$ as a function of the Reynolds number for the same flow.\label{fig_Carleman_Reynolds}}
\end{figure}

In a recent paper, Cheung and coworkers argued that lattice Boltzmann might
be more convenient for Carleman linearization, since the fluid nonlinearity
is formally encoded in the Mach number instead of the Reynolds number, which
makes of course a huge difference. 
They performed a classical Taylor-Green vortex simulation based on a 
Carleman-Lattice Boltzmann scheme, showing excellent agreement at moderate
Mach number, with just three Carleman levels \cite{MARGIE}. 
More recently, in Ref.~\cite{SANAVIO24}, the authors showed the convergence of the 
Carleman method to the analytical solution with the Carleman system truncated at second order 
for fluids at moderate-low Reynolds number, up to $100$ (see Figure \ref{fig_Carleman_Reynolds}). 

% ------------------------------------------------------
% -----------------------------------------------------

\begin{figure}
\centering
\includegraphics[scale=0.2]{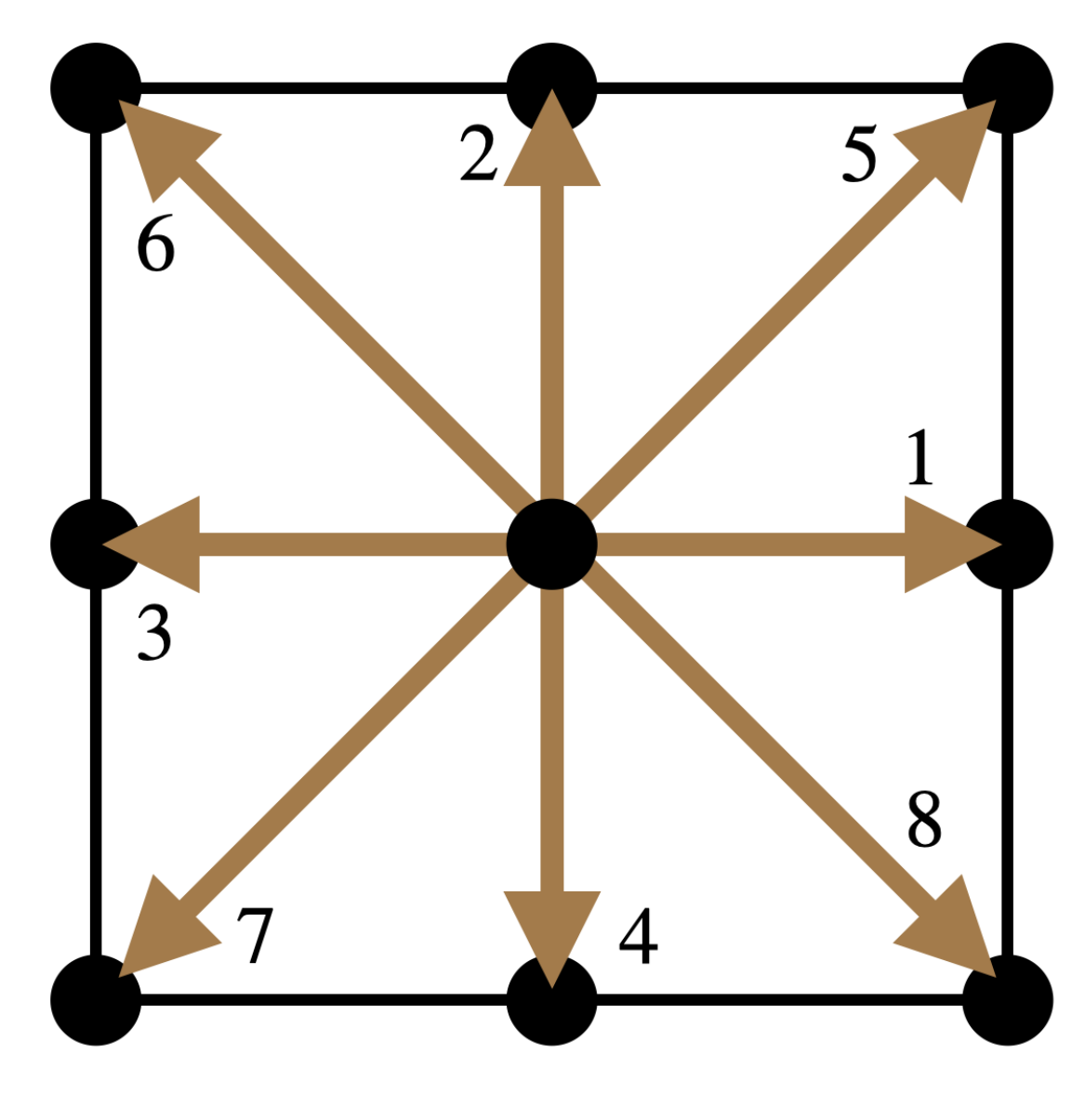}
\caption{The D2Q9 model, defined on a two-dimensional square lattice, with the corresponding set 
of nine velocity vectors labeled by index $i$, cf. Table~\ref{tab_D2Q9}.\label{fig_D2Q9}}
\end{figure}

Let us remind that the Lattice Boltzmann (LB) method is based on 
the following scheme:
In equations:
\begin{equation}
f_i({\bf x} + {\bf v}_i \Delta t, t + \Delta t) -f_i({\bf x},t) 
= - \frac{f_i-f_i^{eq}}{\tau}
\end{equation}
where $f_i({\bf x},t)$, is the probability to find a
representative fluid parcel with discrete velocity ${\bf v}_i$ at position ${\bf x}$
and time $t$. In the above, $f_i^{eq}$ is the corresponding discrete local equilibrium
and $\tau$ is a local relaxation time, fixing the viscosity of the lattice fluid. 
Importantly, the local equilibrium is a quadratic function of 
the Mach number $Ma=u/c_s$, $c_s$ being the speed of sound.
A defining feature of the LB method is that it allows to use a 
fixed, generally small, number of discrete velocities ${\bf v}_i$. 
For instance, it turns out that just nine discrete velocities are sufficient
to fully recover the physical properties of a two-dimensional flow defined on a square lattice. 
The nine velocities are depicted in Fig.~\ref{fig_D2Q9} and described in Table~\ref{tab_D2Q9}. 
The model is called D2Q9.
A key feature of LB, not shared by the macroscopic representation of fluid dynamics, is that
streaming proceeds along straight lines, defined by the discrete velocities.
This operation is {\it exact}, in that it does not involve any floating-point calculation
but just a memory shift from the lattice position ${\bf x}$ to another lattice
position ${\bf x} + {\bf v}_i \Delta t$. 
Also to be observed that collisions are completely local, hence perfectly amenable to
parallel computing. Because of this, dissipation is an emergent property which does
not require any second order spatial derivative, as it is the case for the Navier-Stokes
equations. Besides their mathematical elegance, these properties are at the roots of the
computational efficiency of the method and its amenability to parallel computing.

% --------------------------------------------
% --------------------------------------------
\begin{table}
\begin{center}
\begin{tabular}{l | c | c | c | c | c | c | c | c | c |} 
 \hline
i & $0$ & $1$ & $2$ & $3$& $4$ & $5$ & $6$ & $7$& $8$\\
 \hline 
 \hline 
 $c_i$ &  $(0,0)$ &$(1,0)$&$(0,1)$&$(-1,0)$&$(0,-1)$&$(1,1)$&$(-1,1)$&$(-1,-1)$&$(1,-1)$\\
\hline
\end{tabular}
\end{center}
\caption{The velocity set of the D2Q9 model.\label{tab_D2Q9}}
\end{table}

A key point of using the discrete-velocity Boltzmann formalism instead of Navier-Stokes 
is that, owing to the double dimensional phase-space, in the latter non-locality (streaming) 
is linear while non-linearity (collision) is local, while in Navier-Stokes the two merge into a single 
${\bf u} \nabla {\bf u}$ convective term. 

On classical computers this disentanglement proves extremely
beneficial because of the local nature of the collision term, which makes the system 
highly amenable to parallel computation. 
On quantum computers instead it is not simple to relate together the streaming and the collision steps,
as we are going to show next. 
When writing the Lattice Boltzmann functions $f_i(x,t)$ in a quantum state we 
can employ amplitude encoding, using a  quantum register $|\cdot\rangle_v$ for embedding the
discrete velocities $v_i$ and a second  quantum register for encoding the 
lattice sites $|\cdot\rangle_x$. 
The fluid is then described by the following quantum state
\begin{equation}
|\psi(t)\rangle = \sum_i\sum_x f_i(x,t)|i\rangle |x\rangle.
\end{equation}

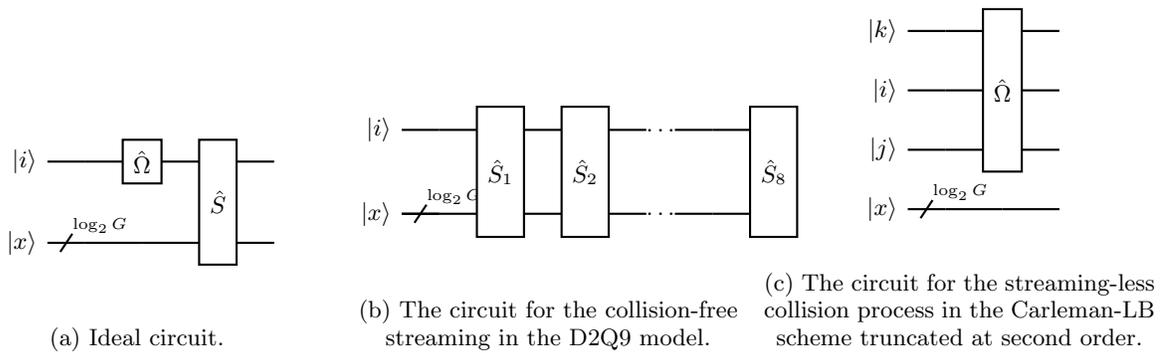
\begin{figure}[t]
\begin{subfigure}{.29\linewidth}
    \centering
\begin{quantikz}
\lstick{$\ket{i}$} &\qw&\gate{\hat{\Omega}}&\gate[2]{\hat{S}}  &\qw\\
\lstick{$\ket{x}$} &\qwbundle{\log_2G}&\qw			&	& \qw\\
\end{quantikz}
\caption{Ideal circuit.}
  \end{subfigure}%
  \hspace{0.5em}
\begin{subfigure}{.29\linewidth}
    \centering
\begin{quantikz}
\lstick{$\ket{i}$} &\qw&\gate[2]{\hat{S}_1}&\gate[2]{\hat{S}_2}&\qw\ldots&\qw&\gate[2]{\hat{S}_8}\\
\lstick{$\ket{x}$} &\qwbundle{\log_2G}\qw&	&   &\qw\ldots&\qw&\\
\end{quantikz}    
\caption{The circuit for the collision-free streaming in the D2Q9 model.}
  \end{subfigure}%
  \hspace{0.5em}
\begin{subfigure}{.29\linewidth}
    \centering
\begin{quantikz}
\lstick{$\ket{k}$} &\qw&\gate[3]{\hat{\Omega}}&\qw\\
\lstick{$\ket{i}$} &\qw&&\qw\\
\lstick{$\ket{j}$} &\qw&&\qw\\
\lstick{$\ket{x}$} &\qwbundle{\log_2G}&\qw&\qw\\
\end{quantikz}    \caption{The circuit for the streaming-less collision process in the Carleman-LB scheme truncated at second order.}
  \end{subfigure}%
  \hspace{0.5em}
\caption{The sketch of three quantum circuits for different Lattice-Boltzmann implementations for a grid with $G$ lattice sites for the D2Q9 model. In (a)   the collision is applied on the velocity register whereas the streaming is a global operation applied on both the position and the velocity registers. In (b) The evolution is obtained as a series of streaming in the 8 velocities of the D2Q9 model. In (c) the quantum register $|k\rangle$ holds the information about the truncation order, whereas $|i\rangle$ and $|j\rangle$ are the velocities associated to the Carleman variables.\label{fig_Circuits}}
\end{figure}

Since the collision step is local and depends only on the velocities of the fluid in the lattice site, one 
may apply an operator $\hat{\Omega}$ which performs the collision step on  the quantum register 
adapted to the velocities and an other operator $\hat{S}$ encoding the streaming process 
to be applied to all quantum registers, being it both non-local and velocity dependent. 
This ideal circuit is shown in Fig.~\ref{fig_Circuits}(a). 
It turns out that since $\hat{\Omega}$ is a non-linear and non-unitary operator, this circuit 
cannot be realized, unless specific symmetry conditions hold, such as those
used in Ref.\cite{YEPEZ1}. 
Unfortunately, these do not conform to the universal quantum computation paradigm.  

We may then opt for hybrid strategies whereby only streaming (collisions) would
be performed on a quantum computer, leaving collisions 
(streaming) to a classical one. 
Interestingly, this leads to different circuits, as sketched in Figure~\ref{fig_Circuits}. 
Fig.~\ref{fig_Circuits}(b) and Fig.~\ref{fig_Circuits}(c) show the circuits for 
collision-free streaming and streaming-free collision processes, that we are going to discuss next. 

\subsubsection{Collision-free streaming}

The quantum computing algorithm for streaming is based on the approach used 
by Steijl and co-workers\cite{RENE2020}.
The streaming is a non-local process shifting the particles in the 
next neighboring site, depending on the corresponding velocity, without
affecting their value. 
Thus, this step can be executed by moving the probability function 
$f_i(\mathbf{x},t)$ to $f_i(\mathbf{x}+\mathbf{v}_i,t+1)$ at the next time step. 

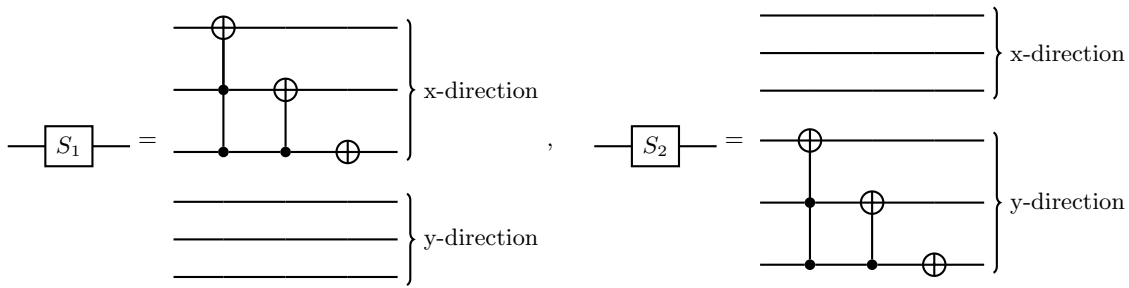
\begin{figure}[t]
\centering
\begin{quantikz}
\qw &\gate{S_1}  &\qw
\end{quantikz}= 
\begin{quantikz}
&\targ{}  & \qw       & \qw    &\rstick[wires=3]{x-direction}\qw\\
&\ctrl{-1}& \targ{}   & \qw    &\qw\\
&\ctrl{-2}& \ctrl{-1} & \targ{}&\qw\\
&\qw      &  \qw      & \qw    &\rstick[wires=3]{y-direction}\qw\\
&\qw      &  \qw      & \qw    &\qw\\
&\qw      &  \qw      & \qw    &\qw
\end{quantikz},
\quad \begin{quantikz}
\qw &\gate{S_2}  &\qw
\end{quantikz}= 
\begin{quantikz}
&\qw      &  \qw      & \qw    &\rstick[wires=3]{x-direction}\qw\\
&\qw      &  \qw      & \qw    &\qw\\
&\qw      &  \qw      & \qw    &\qw\\
&\targ{}  & \qw       & \qw    &\rstick[wires=3]{y-direction}\qw\\
&\ctrl{-1}& \targ{}   & \qw    &\qw\\
&\ctrl{-2}& \ctrl{-1} & \targ{}&\qw
\end{quantikz}
\caption{The quantum circuits for the operators $S_1$ and $S_2$, cf. with Table~\ref{tab_D2Q9}, for a D2Q9 model define on a $8\times8$ lattice. The first three qubits are used to encode the position in the $x$ direction and the latter three for the position in the $y$ direction. They are applied to the quantum state only if the velocity register has value $1$ or $2$ respectively.  \label{fig_Streaming}}
\end{figure}

The corresponding operation on the quantum state reads as follows:

\begin{equation}
|\psi(t)\rangle = \sum_i\sum_{\mathbf{x}}f_i(\mathbf{x},t)|i\rangle|\mathbf{x}
\rangle\xrightarrow{\hat{S}}|\psi(t+1)\rangle =  
\sum_i\sum_{\mathbf{x}}f_i(\mathbf{x},t)|i\rangle|\mathbf{x}+\mathbf{v}_i\rangle
\end{equation}

\noindent i.e. only the positions register is affected by the streaming process, 
whereas the velocity register acts as control. 
The streaming operator is obtained as a combination of the streaming in the different directions, for the D2Q9 model with 8 velocities (plus the null velocity) is $\hat{S} = \bigoplus_{i=1}^8\hat{S}_i,$
that are therefore applied depending on the value of the qubit in the velocity register. The streaming can be obtained by applying a series of multiqubit controlled operations on the position register. We show in Figure~\ref{fig_Streaming} an example for the $\hat{S}_1$ streaming operator when applied on a $8\times 8$ lattice.

\subsubsection{Streaming-free collision}
A completely different approach is employed to perform the collision step: Carleman 
linearization.
Based on the Lattice Boltzmann method, Itani et al.\cite{WAEL} developed an 
approach termed \textit{Carleman for second-quantized Lattice Boltzmann}. 
This terminology stems from the fact that the Boltzmann operator, defined by:
\begin{equation}
\mathcal{B} f_i = - {\bf v}_i \cdot \nabla f_i - \frac{f_i-f_i^{eq}}{\tau},
\end{equation}
can be expressed in terms of the second-quantization annihilation and generation 
operators via the relation $\nabla = (\hat a-\hat a^+)$. Collisions follow the 
bosonic encoding first proposed by Mezzacapo et al. \cite{MEZZA}, whereby dissipative
effects are represented as a weighted sum of two unitary operators. 
The scheme is as "Feynmansque" as it can possibly get, since it builds on 
a one-to-one analogy between LB and the Dirac equation, as first proposed in \cite{QLB}.
As a result, the formal solution $f_t = e^{\mathcal{B}t} f_0$ can be 
computed in analogy with quantum mechanics.
However, it is  subject to a number of additional questions: primarily 
truncation effects due to the finite number of bosonic excitations and 
the long-time behaviour of non-unitarity errors \cite{WAEL23}. 

On a similar but operationally different line, Sanavio et al.~\cite{SANAVIO24} developed 
a quantum circuit implementing the collision step of a Carleman-Lattice Boltzmann algorithm.
Since the number of variables with the Carleman method increases exponentially 
with the truncation order $k$, the circuit needs 
more qubits to embed the relevant information. 
The collision step requires $k+1$ additional quantum registers, embedding 
the truncation order and the products between local Boltzmann functions with different 
velocities, say $f_i(x), f_i(x)f_j(y), f_i(x)f_j(y)f_k(z)$ and so on.
With this setting, the collision quantum circuit is genuinely local, as it depends 
only on the velocity registers and it  can be implemented using only a fixed number of 
two-qubit gates, regardless of the number of lattice sites. 
This meets the promise of quantum advantage but unfortunately it is 
inconsistent with the streaming process, which must therefore
be directed to a classical machine. 
This may seem an ideal option for hybrid quantum computing, since streaming is a 
floating-point free operation. However, accessing data is by no means 
a cost-free operation and it is indeed known to take a significant
fraction of the computer time spent on collisions \cite{EXALB}.

\subsubsection{Fully quantum algorithm: Collision and Streaming}

In Ref.~\cite{SANAVIO24}, the authors managed to generalize the circuit for the streaming step to Carleman 
variables of higher orders, thus allowing the subsequent application of 
streaming and collision step on Carleman variables with a single embedding 
and a single readout of the results after evolution.
This results in a fully quantum algorithm.
In order to embed different non-linear terms one needs $2(k+1)$ additional quantum registers, one 
encoding the information about the degree of non linearity up 
to order $k$, and the other $2k$ to embed position and velocity of 
the multiproduct functions.  

This method was found to meet with an exponential 
depth of the quantum circuit as a function of the number of qubits.
The reason is that the collision operator cannot be written in the form
of a sparse and combinatorially block-diagonal matrix, nor in terms of 
a compact combination of Pauli matrices.
Hence, to date, it cannot be quantum compiled.  

Possible ways out are currently under investigation: one option is the application of the 
Carleman linearization method directly to the Navier-Stokes equations. 
The advantage of the Carleman-Navier-Stokes procedure rests a much lesser number of
Carleman variables, with a corresponding reduction of the circuit depth.
The flip side, however, is a more complex structure of the Carleman 
matrix, which, combined with the Reynolds versus Mach number as a measure
of nonlinearity, may impair Carleman convergence.
A detailed analysis along this direction is currently underway.

\section{Conclusions}

In summary, we have presented a survey of the a few leading approaches to 
the quantum simulation of classical fluids. 
Various obstacles stand in the way of the efficient simulation of fluid flows
on quantum computers, especially at high-Reynolds numbers, mostly on 
account of the strong nonlinear effects.
A few ways potential ways out have been illustrated, but their practical 
implementation commands major advances both on the algorithmic side,
let alone quantum technology \cite{DAS}.  

Before closing, it should be observed that while, everybody would
expect quantum computers to handle turbulence, the physics of fluids 
is littered  with interesting  problems at low-Reynolds, especially in 
microfluidics, soft matter and biology 
\cite{STONE,MICRO,OUP2}, but also in high-energy physics, typically 
quark-gluon plasma experiments.
Hence, even in case turbulent flows would prove beyond reach 
to quantum computing, there would nevertheless be plenty of 
interesting fluid applications in the low-moderate Reynolds regime.

For instance, it could be of great interest to devise a 
{\it quantum multi-scale} application, coupling quantum algorithms for 
biomolecules in a water solvent described by a quantum algorithm for
low-Reynolds fluid flow. Given that low-Reynolds flows are nonlocal, perhaps
the inherent nonlocality of quantum mechanics could prove helpful in
representing the classical nonlocality of low-Reynolds flows.

On a more philosophical ground, should the physics of fluids show a case for
"classical advantage", there would still be interesting lessons to learn.
Indeed, as observed earlier on, while it is true that Nature isn't classical, it 
is equally true that Nature has a very strong tendency to {\it become}
classical at macroscopic scales/high temperatures. 
Withstanding such tendency, or perhaps even take of advantage of it,
is indeed the prime struggle of quantum computers. 
Looking into ways to achieve this very hard goal is not only of 
practical but also of major foundational interest, since  it is likely to
shed new light into the still open problem of "classicalization", i.e. the emergence
of classical behaviour out of underlying quantum mechanical microscopic physics. 
This is only one (outstanding) example of the fact that
quantum computers offer the unique chance to ask questions that
the founding fathers of quantum mechanics could only formulate 
as "Gedanken Experiments".

\acknowledgments

The authors have benefited from valuable discussions 
with many colleagues, particularly
P. Coveney, N. Defenu, 
G. Galli, M. Grossi, B. Huang, W. Itani, A. Mezzacapo,  
S. Ruffo, A. Solfanelli K. Sreenivasan, R. Steijl, and T. Weaving.
The authors also acknowledge financial support form the Italian National
Centre for HPC, Big Data and Quantum Computing (CN00000013).

\end{document}